\PassOptionsToPackage{unicode}{hyperref}
\PassOptionsToPackage{hyphens}{url}
\documentclass[
  12pt,
  a4paper]{article}
\usepackage{amsmath,amssymb}
\usepackage{lmodern}
\usepackage{setspace}
\usepackage{ifxetex,ifluatex}
\ifnum 0\ifxetex 1\fi\ifluatex 1\fi=0 % if pdftex
  \usepackage[T1]{fontenc}
  \usepackage[utf8]{inputenc}
  \usepackage{textcomp} % provide euro and other symbols
\else % if luatex or xetex
  \usepackage{unicode-math}
  \defaultfontfeatures{Scale=MatchLowercase}
  \defaultfontfeatures[\rmfamily]{Ligatures=TeX,Scale=1}
\fi
\IfFileExists{upquote.sty}{\usepackage{upquote}}{}
\IfFileExists{microtype.sty}{% use microtype if available
  \usepackage[]{microtype}
  \UseMicrotypeSet[protrusion]{basicmath} % disable protrusion for tt fonts
}{}
\makeatletter
\@ifundefined{KOMAClassName}{% if non-KOMA class
  \IfFileExists{parskip.sty}{%
    \usepackage{parskip}
  }{% else
    \setlength{\parindent}{0pt}
    \setlength{\parskip}{6pt plus 2pt minus 1pt}}
}{% if KOMA class
  \KOMAoptions{parskip=half}}
\makeatother
\usepackage{xcolor}
\IfFileExists{xurl.sty}{\usepackage{xurl}}{} % add URL line breaks if available
\IfFileExists{bookmark.sty}{\usepackage{bookmark}}{\usepackage{hyperref}}
\hypersetup{
  pdftitle={Tinjauan atas Efektivitas Penggunaan Key Opinion Leader (KOL) dalam Penjualan Surat Utang Negara Ritel seri SBR011},
  pdfauthor={Dea Avega Editya},
  hidelinks,
  pdfcreator={LaTeX via pandoc}}
\usepackage[left=3cm,right=3cm,top=2cm,bottom=2cm]{geometry}
\usepackage{longtable,booktabs,array}
\usepackage{calc} % for calculating minipage widths
\usepackage{etoolbox}
\makeatletter
\patchcmd\longtable{\par}{\if@noskipsec\mbox{}\fi\par}{}{}
\makeatother
\IfFileExists{footnotehyper.sty}{\usepackage{footnotehyper}}{\usepackage{footnote}}
\makesavenoteenv{longtable}
\usepackage{graphicx}
\makeatletter
\def\maxwidth{\ifdim\Gin@nat@width>\linewidth\linewidth\else\Gin@nat@width\fi}
\def\maxheight{\ifdim\Gin@nat@height>\textheight\textheight\else\Gin@nat@height\fi}
\makeatother
\setkeys{Gin}{width=\maxwidth,height=\maxheight,keepaspectratio}
\makeatletter
\def\fps@figure{htbp}
\makeatother
\usepackage{float}
\usepackage{booktabs}
\usepackage{longtable}
\usepackage{array}
\usepackage{multirow}
\usepackage{wrapfig}
\usepackage{float}
\usepackage{colortbl}
\usepackage{pdflscape}
\usepackage{tabu}
\usepackage{threeparttable}
\usepackage{threeparttablex}
\usepackage[normalem]{ulem}
\usepackage{makecell}
\usepackage{xcolor}
\ifluatex
  \usepackage{selnolig}  % disable illegal ligatures
\fi
\usepackage[]{biblatex}
\title{Tinjauan atas Efektivitas Penggunaan Key Opinion Leader (KOL) dalam Penjualan Surat Utang Negara Ritel seri SBR011}
\usepackage{etoolbox}
\makeatletter
\providecommand{\subtitle}[1]{% add subtitle to \maketitle
  \apptocmd{\@title}{\par {\large #1 \par}}{}{}
}
\makeatother
\subtitle{dengan pendekatan data analytics}
\author{Dea Avega Editya}
\date{6/17/2022}

\begin{document}
\maketitle
\begin{abstract}
Indonesian Ministry of Finance had endorsed 10 Key Opinion Leaders to help promoting government retail bonds SBR011 during selling period of 25 May-16 June 2022. This study analyzed effectiveness of the endorsement by using several indicators; engagement rate, enthusiasm rate and sentiment analysis of feedbacks from KOL audiens. Data was gathered from social media Instagram and TikTok social platform used by the KOL to post their marketing contents. This paper found that the endorsement is quite effective to promote the SBR011 and yields mostly positive feedback on the marketing campaign.
\end{abstract}

\setstretch{1.5}
\hypertarget{latar-belakang}{%
\subsection{Latar Belakang}\label{latar-belakang}}

Paper ini disusun untuk melakukan tinjauan umum atas kegiatan pemasaran produk investasi Savings Bonds Ritel (SBR) seri SBR011 yang dilakukan oleh \textbf{Key Opinion Leaders} (KOL). Proses analisis dalam paper ini menggunakan software R beserta beberapa packages yang diperlukan.

Hasil tinjauan atas kegiatan KOL ini diharapkan dapat memberikan gambaran mengenai efektivitas KOL dalam membantu Pemerintah memasarkan dan terutama mengedukasi masyarakat mengenai produk SBR011. Dalam melakukan tinjauan terhadap efektivitas KOL, penulis menggunakan inidkator \emph{proxy} berupa tingkat \emph{engagement}, antusiasme serta sentimen umum dari audiens KOL tersebut, mengingat nilai penjualan SBR011 yang dihasilkan dari penggunaan jasa KOL tersebut tidak dapat diperoleh.

Seluruh \emph{workflow} dalam paper ini dapat direproduksi kembali oleh peneliti, akademisi maupun mahasiswa yang memiliki minat terhadap bidang analisis data terutama yang berkaitan dengan produk investasi surat berharga negara.

\hypertarget{kegiatan-pemasaran-sbr011}{%
\subsubsection{Kegiatan Pemasaran SBR011}\label{kegiatan-pemasaran-sbr011}}

Surat Utang Negara (SUN) Ritel berjenis Saving Bonds kembali ditawarkan kepada masyarakat warga negara Indonesia oleh Pemerintah melalui Direktorat Surat Utang Negara (Dit.SUN) Kementerian Keuangan. Pada periode 25 Mei-16 Juni 2022, masyarakat dapat melakukan pemesanan untuk instrumen Saving Bonds seri SBR011. Seperti halnya pada penerbitan SUN ritel sebelumnya (ORI021) pada bulan Maret lalu, pemesanan SBR011 dapat dilakukan melalui 28 Mitra Distribusi yang telah mendapatkan otorisasi dari Pemerintah untuk membantu penjualan instrumen obligasi tersebut (daftar lengkap Midis SBR011 dapat dilihat pada \emph{landing page} Kemenkeu: \url{https://www.kemenkeu.go.id/single-page/savings-bond-ritel/}).

Dalam melakukan pemasaran SBR011, Dit. SUN memberikan dua opsi metode yang dapat dipilih oleh Midis sesuai preferensinya masing-masing. Metode pertama adalah menggunakan kegiatan berbentuk sosialisasi, edukasi maupun kegiatan yang sifatnya lebih terbatas kepada kalangan internal Midis semisal temu nasabah.
Metode kedua yang digunakan dan baru pertama kali diterapkan pada penerbitan SBR011 yaitu menggunakan Key Opinion Leader (KOL). KOL mengandalkan tokoh publik yang dipandang memiliki kekuatan basis massa yang ditandai dengan jumlah follower yang besar. Penggunaan KOL bertujuan untuk membangun \emph{``trusted relationship''} khususnya terhadap instrumen SBR011 yang sedang ditawarkan oleh Pemerintah (sumber: dokumen internal Direktorat Surat Utang Negara Kemenkeu).

Jumlah mitra distribusi yang melaksanakan kegiatan sosialisasi virtual sebanyak 20, adapun midis yang melaksanakan marketing melalui konten Key Opinion Leader (KOL) sebanyak 8 midis. Total audiens sebanyak 24901 yang terdiri dari 2904 jumlah peserta sosialisasi virtual dan 21997 jumlah like dari konten KOL.

\hypertarget{definisi-key-opinion-leader-kol}{%
\subsubsection{Definisi Key Opinion Leader (KOL)}\label{definisi-key-opinion-leader-kol}}

Menurut influencermarketinghub, KOL dideskripsikan sebagai ``\emph{person or organization who has expert product knowledge and influence in a respective field. They are trusted by relevant interest groups and have significant effects on consumer behavior}'' \autocite{Geyser}. Lebih lanjut, KOL dapat dianggap sama dengan \emph{influencer}, kecuali bahwa KOL dapat menggunakan media \emph{online} maupun \emph{offline} sedangkan influencer umumnya menggunakan media \emph{online} saja. Untuk tujuan penulisan artikel ini, kata KOL dan influencer akan dianggap memiliki makna yang sama sehingga digunakan secara bergantian.

Dalam rangka membantu Midis dalam memilih KOL untuk pemasaran SBR011, Dit. SUN merekomendasikan empat tipe KOL yang dapat di-\emph{endorse} yaitu \textbf{Student, Professional, Enterpreneur dan Housewives}. Untuk tipe Student, KOL merupakan figur yang digemari kalangan anak muda dan berstatus sebagai pelajar dan mahasiswa. Untuk tipe Professional, figur merupakan pekerja profesional yang berjiwa muda dan biasa membagikan konten seputar profesinya. Tipe Enterpreneur merupakan kelompok KOL yang memiliki bisnis, berjiwa muda dan membagikan kisah inspiratif seputar bisnis yang digeluti. Adapun untuk tipe Housewive, KOL adalah para wanita yang menikmati perannya sebagai ibu rumah tangga dan cukup aktif membagikan kegiatannya sehari-hari. Seluruh tipe tersebut tentunya harus \emph{concern} terhadap dunia investasi, memiliki sentimen positif yang menginspirasi serta tidak pernah mengunggah konten yang berbau SARA, pornografi maupun pornoaksi.

Berdasarkan jumlah followernya, KOL dapat dibagi menjadi lima kelompok yaitu Nano (memiliki rentang jumlah follower 1000-10.000), Micro (10.000-50.000), Mid-Tier (50.000-500.000), Macro (500.000-1 juta), dan Mega (\textgreater{} 1 juta) \autocite[sumber: mediakix.com sebagaimana dikutip dalam][]{Geyser}.

\hypertarget{penggunaan-kol-dalam-marketing-sbr011}{%
\subsubsection{Penggunaan KOL dalam Marketing SBR011}\label{penggunaan-kol-dalam-marketing-sbr011}}

Untuk keperluan marketing SBR011, Midis dapat memilih KOL yang berasal minimal dari kategori Mikro yang memiliki \emph{engagement rate} tinggi. Hal ini sejalan dengan argumen dari Geyser bahwa influencer online yang dianggap sukses biasanya merupakan kategori Micro dan Mid-Tier \autocite*{Geyser}. Terdapat hal lain yang dipersyaratkan dalam panduan pemilihan KOL yaitu influencer harus memiliki karakteristik yang mendekati target investor SBR011 serta memiliki \emph{engaged-followers} yang berkaitan dengan konten seputar investasi. Konten yang diunggah pun harus dapat menyampaikan pesan gerakan kolektif untuk membantu pemulihan ekonomi dan pembangunan nasional. Sementara gaya penyampaian yang direkomendasikan ialah \emph{storytelling} dan persuasif dalam menjelaskan detail proses dan konsisten memasukkan unsur \emph{value} dalam kontennya.

\hypertarget{influencer-kol-sbr011}{%
\subsubsection{Influencer KOL SBR011}\label{influencer-kol-sbr011}}

Berdasarkan kriteria yang ditetapkan tersebut, terdapat sepuluh (10) KOL yang di-\emph{endorse} oleh Midis untuk membantu melakukan kampanye marketing SBR011 selama periode penawaran. Dalam melakukan tinjauan atas pelaksanaan KOL, penulis mengumpulkan data profil dari masing-masing KOL dengan informasi yang diperoleh dari platform analisa.io \autocite{Analisa}. Informasi profil tersebut merupakan \emph{snapshot} data per tanggal 20 Juni 2022 yang meliputi; jumlah follower, jumlah post, jumlah like, rata-rata like per post yang diunggah.

Data tersebut kemudian dilengkapi dengan tema konten dan tipe audiens yang ditentukan oleh penulis setelah melakukan observasi terhadap unggahan tiap KOL. Secara garis besar, tema unggahan konten KOL terbagi menjadi dua yaitu keuangan dan umum, adapun tipe audiens oleh penulis dibagi menjadi muda dan sangat muda.

Seluruh \emph{influencer} yang menggunakan platform TikTok dalam kampanye SBR011 (Fayza, Chornella dan Felicia) dianggap memiliki audiens yang berusia sangat muda, sesuai dengan statistik pengguna platform tersebut yang mayoritas berada di rentang usia 13-24 tahun \autocite{Dsouza}. Adapun Morgan Oey merupakan satu-satunya influencer berplatform instagram yang diasosiasikan oleh penulis dengan audiens berusia sangat muda mengingat sosok yang merupakan mantan vokalis band Smash tersebut cukup populer dan digandrungi oleh remaja.

Pada akhirnya, penulis juga menambahkan data hasil observasi jumlah like pada konten SBR011 yang diunggah tiap KOL untuk mengukur antusiasme follower terhadap produk SBR011 yang ditawarkan. Profil KOL SBR011 secara umum ditampilkan pada tabel \ref{tab:table}:

\begin{table}[H]

\caption{\label{tab:table}Profil KOL SBR011}
\centering
\fontsize{9}{11}\selectfont
\begin{tabular}[t]{l|l|l|l|l}
\hline
\cellcolor{green}{KOL} & \cellcolor{green}{Tipe} & \cellcolor{green}{Jumlah Follower} & \cellcolor{green}{Tema} & \cellcolor{green}{Audiens}\\
\hline
Melvin Mumpuni & Professional & Mid-Tier (50k - 500k) & keuangan & muda\\
\hline
Lolita Setyawati & Professional & Micro (10k - 50k) & keuangan & muda\\
\hline
Samuel Ray & Professional & Mid-Tier (50k - 500k) & keuangan & muda\\
\hline
Vina Muliana & Professional & Mid-Tier (50k - 500k) & keuangan & muda\\
\hline
Sigiwimala & Professional & Mid-Tier (50k - 500k) & umum & muda\\
\hline
Dewi Andarini & Professional & Micro (10k - 50k) & keuangan & muda\\
\hline
Morgan Oey & Professional & Mid-Tier (50k - 500k) & umum & sangat muda\\
\hline
Fayza & Student & Mid-Tier (50k - 500k) & umum & sangat muda\\
\hline
Chornella & Professional & Mid-Tier (50k - 500k) & keuangan & sangat muda\\
\hline
Felicia & Student & Mid-Tier (50k - 500k) & keuangan & sangat muda\\
\hline
\end{tabular}
\end{table}

Sejalan dengan tujuan dari kegiatan marketing SBR011 melalui KOL, Pemerintah ingin menjangkau audiens yang lebih luas melalui bantuan dari influencer yang di-endorse oleh Midis. Oleh karenanya, influencer yang dipilih merupakan figur yang dianggap memiliki tingkat engagement tinggi. Menurut sebuah artikel dari www.sproutsocial.com, tingkat engagement dapat dijelaskan sebagai suatu metrik yang digunakan untuk mengukur seberapa aktif follower terhadap konten yang diunggah \autocite{Sprout}. Tingkat \emph{engagement} dapat membantu sebuah perusahaan dalam memperkirakan besaran Return on Investment (ROI) dalam sebuah kampanye \emph{digital advertising} yang dilakukan perusahaan tersebut \autocite*{Sprout}. Tingginya tingkat engagement dapat membuat suatu produk serta brand yang diiklankan menjadi lebih dikenal, memicu iklan dari mulut ke mulut yang lebih masif, meningkatkan citra serta meningkatkan kualitas hubungan dengan pengguna produk \autocite{Sprout}.

Adapun faktor-faktor yang umumnya diperhitungkan dalam perhitungan tingkat \emph{engagement} pada sebuah konten media sosial, semisal Instagram, yaitu jumlah like dan komentar pada konten \autocite*{Sprout}. Sejalan dengan penjelasan tersebut, paper ini menggunakan \textbf{angka rata-rata jumlah like per post} dari KOL untuk mengetahui tingkat \emph{engagement} secara umum dari masing-masing KOL.

\begin{figure}[H]

{\centering \includegraphics{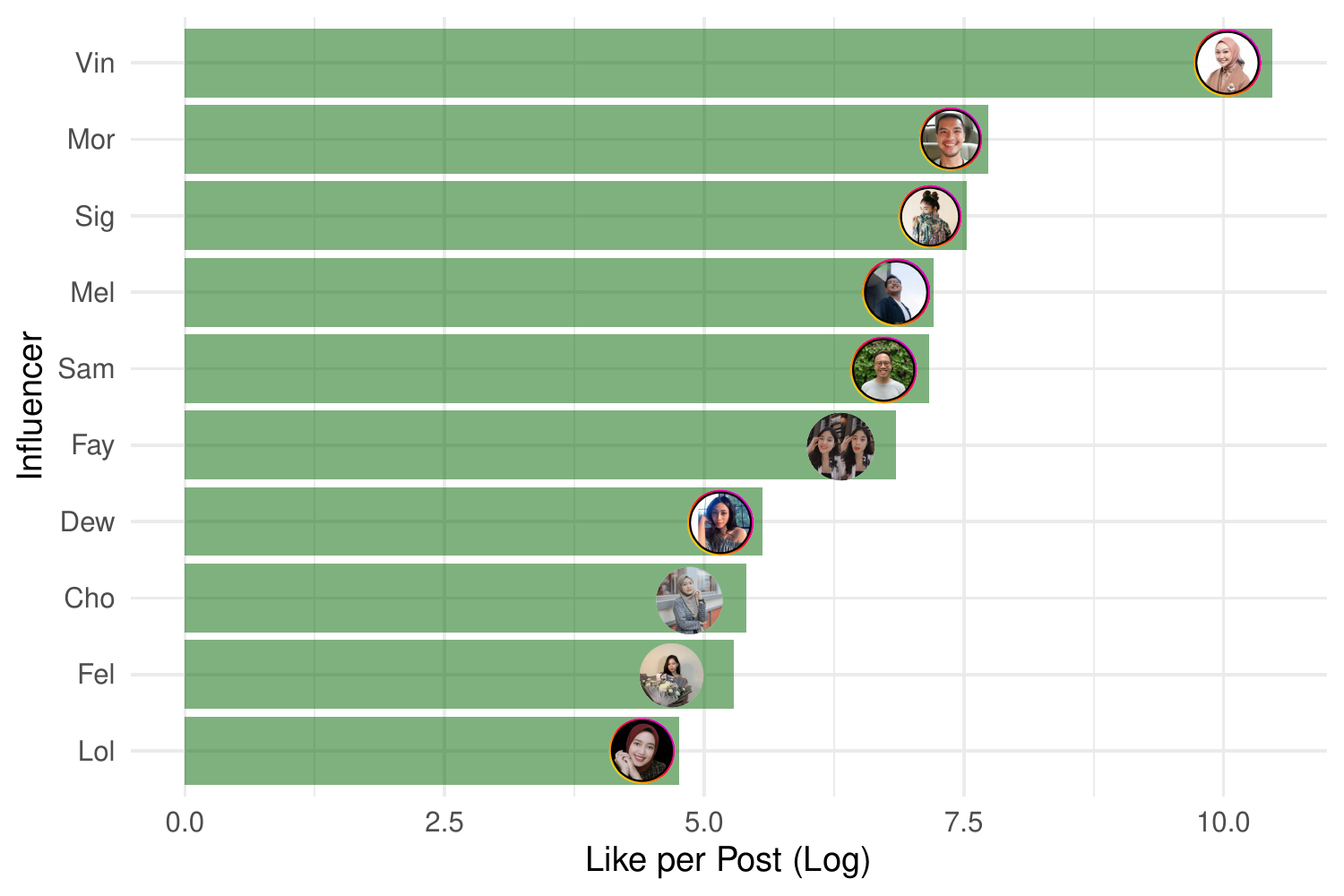} 

}

\caption{Rata-rata Like per Post KOL}\label{fig:liked}
\end{figure}

Berdasarkan data yang dihimpun, seluruh KOL yang di-\emph{endorse} dapat dianggap memiliki tingkat \emph{engagement} yang tinggi karena memiliki rata-rata \emph{like per post} lebih dari 100. Hal ini telah sesuai dengan ketentuan dalam panduan pemilihan KOL yang mensyaratkan tingkat \emph{engagement} KOL yang tinggi. Lebih lanjut, gambaran umum perbandingan tingkat \emph{engagement} KOL ditampilkan pada grafik \ref{fig:liked}.

Grafik \ref{fig:liked} menggunakan satuan logaritma sebab jumlah rata-rata \emph{like per post} salah satu KOL yaitu Vina Muliana sangat ekstrim tingginya jika dibandingkan dengan KOL lain. Setelah ditelusuri lebih jauh, Vina Muliana merupakan salah satu \emph{influencer} papan atas yang masuk dalam daftar Forbes 30 Under 30 untuk kategori Media, Marketing dan Advertising.

\hypertarget{pengelompokan-kol-berdasarkan-kesamaan-karakteristik}{%
\subsubsection{Pengelompokan KOL Berdasarkan Kesamaan Karakteristik}\label{pengelompokan-kol-berdasarkan-kesamaan-karakteristik}}

Para influencer yang jasanya dipakai untuk kampanya penjualan SBR011 dapat dikelompokkan berdasarkan kesamaan karakteristiknya. Metode yang digunakan untuk melakukan pengelompokan menggunakan Principal Component Analysis (PCA) dengan meringkas multivariabel yang menjelaskan karakteristik masing-masing KOL. Terdapat enam variabel yang digunakan dalam proses analisis PCA ini, yaitu \textbf{tipe audiens, tema, jumlah follower, jumlah post, tipe post (video dan gambar) dan rata-rata like per post} (tingkat \emph{engagement}).

\begin{figure}[H]

{\centering \includegraphics{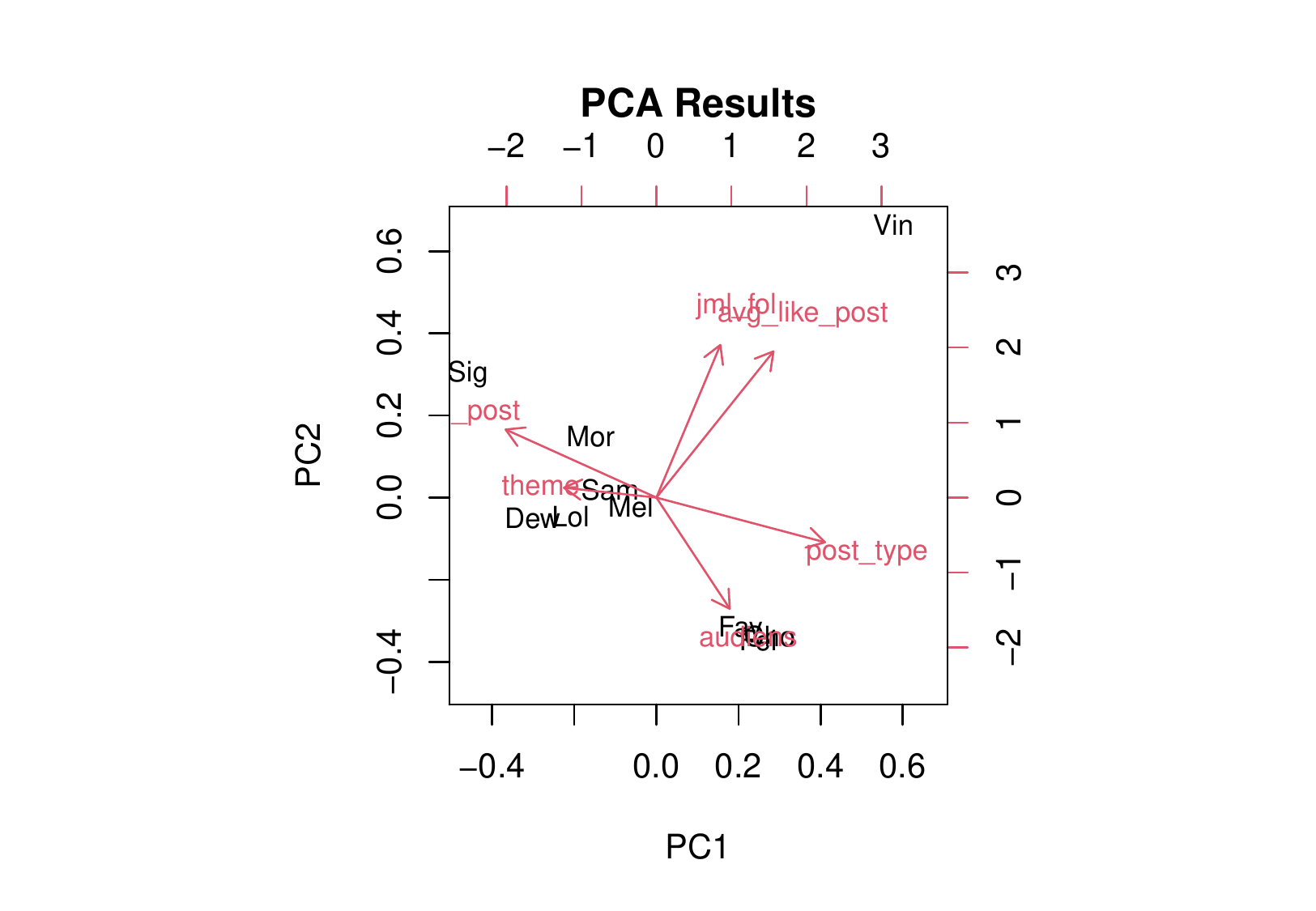} 

}

\caption{Pengelompokan KOL berdasarkan PCA}\label{fig:pca}
\end{figure}

\begin{table}[H]

\centering
\fontsize{9}{11}\selectfont
\begin{tabular}[t]{l|r|r|r|r|r|r}
\hline
\cellcolor{green}{ } & \cellcolor{green}{PC1} & \cellcolor{green}{PC2} & \cellcolor{green}{PC3} & \cellcolor{green}{PC4} & \cellcolor{green}{PC5} & \cellcolor{green}{PC6}\\
\hline
audiens & 0.2549537 & -0.4404453 & 0.5709359 & -0.4099268 & 0.2521745 & -0.4282603\\
\hline
theme & -0.3212526 & 0.0390470 & 0.6689161 & 0.6681646 & -0.0371263 & -0.0010631\\
\hline
jml\_fol & 0.2221315 & 0.6043752 & 0.3683144 & -0.2757917 & 0.3722542 & 0.4848696\\
\hline
jml\_post & -0.5236835 & 0.2695429 & 0.2211074 & -0.5203871 & -0.5631239 & -0.1276811\\
\hline
post\_type & 0.5855714 & -0.1770339 & 0.2017606 & 0.0527748 & -0.6781622 & 0.3498122\\
\hline
avg\_like\_post & 0.4065891 & 0.5789815 & -0.0364997 & 0.1893863 & -0.1394416 & -0.6654490\\
\hline
\end{tabular}
\end{table}

Berdasarkan grafik \ref{fig:pca} yang menggunakan komponen PC1 dan PC2, secara umum KOL SBR011 dapat terbagi menjadi tiga kelompok. Kelompok pertama hanya terdiri dari Vina Muliana, dengan karakteristik tingkat \emph{engagement}, jumlah \emph{follower} dan jumlah unggahan (post) yang besar.
Kelompok kedua terdiri dari 6 KOL yang memiliki karakteristik hampir serupa, yaitu Melvin, Lolita, Morgan, Samuel, Dewi dan Sigi. Keenam influencer tersebut secara umum memiliki kesamaan dari jumlah unggahan konten (post) yang besar serta tema konten yang biasa diunggah.
Kelompok ketiga diwakili oleh \emph{influencer} yang memiliki kesamaan dari sisi tipe audiens yaitu kelompok usia muda, serta antusias \emph{follower} terhadap konten SBR011 yang cukup tinggi, terutama Chornella. Ketiga KOL ini juga menggunakan jenis konten yang serupa yaitu video dalam mengunggah konten SBR011.

\hypertarget{analisis-efektivitas-penggunaan-kol-dalam-penjualan-sbr011}{%
\subsection{Analisis Efektivitas Penggunaan KOL dalam Penjualan SBR011}\label{analisis-efektivitas-penggunaan-kol-dalam-penjualan-sbr011}}

Penilaian efektivitas penggunaan KOL dalam membantu kampanye penjualan SBR011 akan ditinjau melalui tiga indikator yaitu \emph{engagement rate}, antusiasme serta sentimen audiens.

\hypertarget{tingkat-engagement-audiens-atas-konten-sbr011}{%
\subsubsection{\texorpdfstring{Tingkat \emph{Engagement} Audiens atas Konten SBR011}{Tingkat Engagement Audiens atas Konten SBR011}}\label{tingkat-engagement-audiens-atas-konten-sbr011}}

\begin{figure}[H]

{\centering \includegraphics{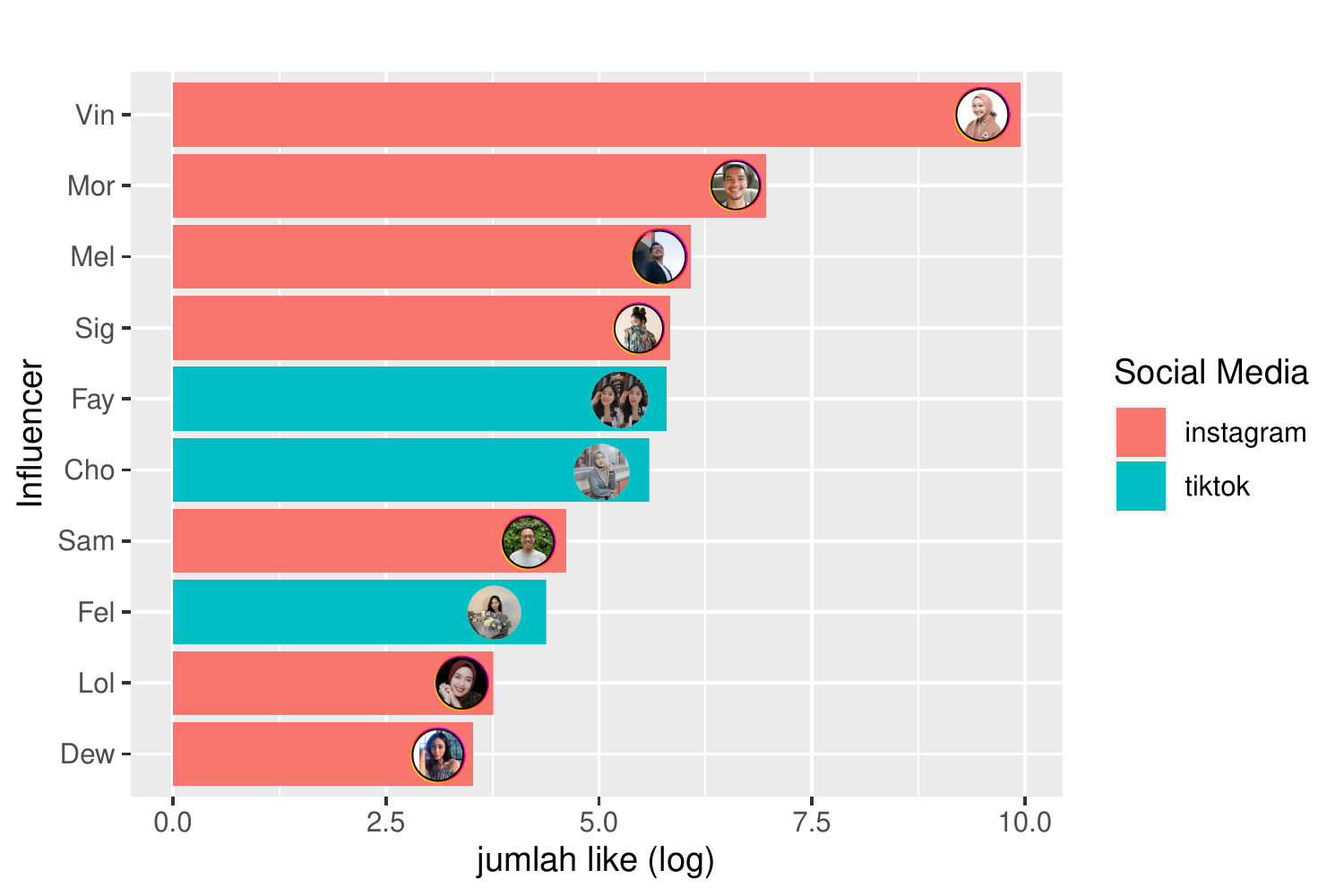} 

}

\caption{Tingkat Engagement terhadap Unggahan Konten SBR011}\label{fig:engage}
\end{figure}

Berdasarkan grafik \ref{fig:engage}, dapat dilihat bahwa konten SBR011 yang diunggah oleh Vina Muliana merupakan konten yang paling menyedot jumlah like yang menjadi indikator tingginya tingkat \emph{engagement} dari followernya. Hal ini tidaklah mengejutkan mengingat KOL tersebut juga memiliki jumlah \emph{follower} paling banyak dibandingkan KOL SBR011 lainnya. Posisi kedua dan ketiga diisi oleh Morgan Oey dan Melvin.

Perbandingan yang ditampilkan pada grafik \ref{fig:engage} kembali menggunakan skala log agar tingkat \emph{engagement} masing-masing KOL dapat lebih mudah diperbandingkan. Apabila menggunakan skala asli, jumlah \emph{engagement (like)} konten SBR011 yang diperoleh oleh Vina Muliana tampak sangat superior dibandingkan jumlah keseluruhan \emph{engagement} yang dikumpulkan oleh KOL lainnya. Hal tersebut menandakan bahwa kampanye SBR011 sejatinya sudah dapat menjangkau audiens secara luas hanya dengan kontribusi dari 1 KOL yaitu Vina Muliana.

\hypertarget{antusiasme-audiens-atas-konten-sbr011}{%
\subsubsection{\texorpdfstring{\emph{Antusiasme} Audiens atas Konten SBR011}{Antusiasme Audiens atas Konten SBR011}}\label{antusiasme-audiens-atas-konten-sbr011}}

\begin{figure}[H]

{\centering \includegraphics{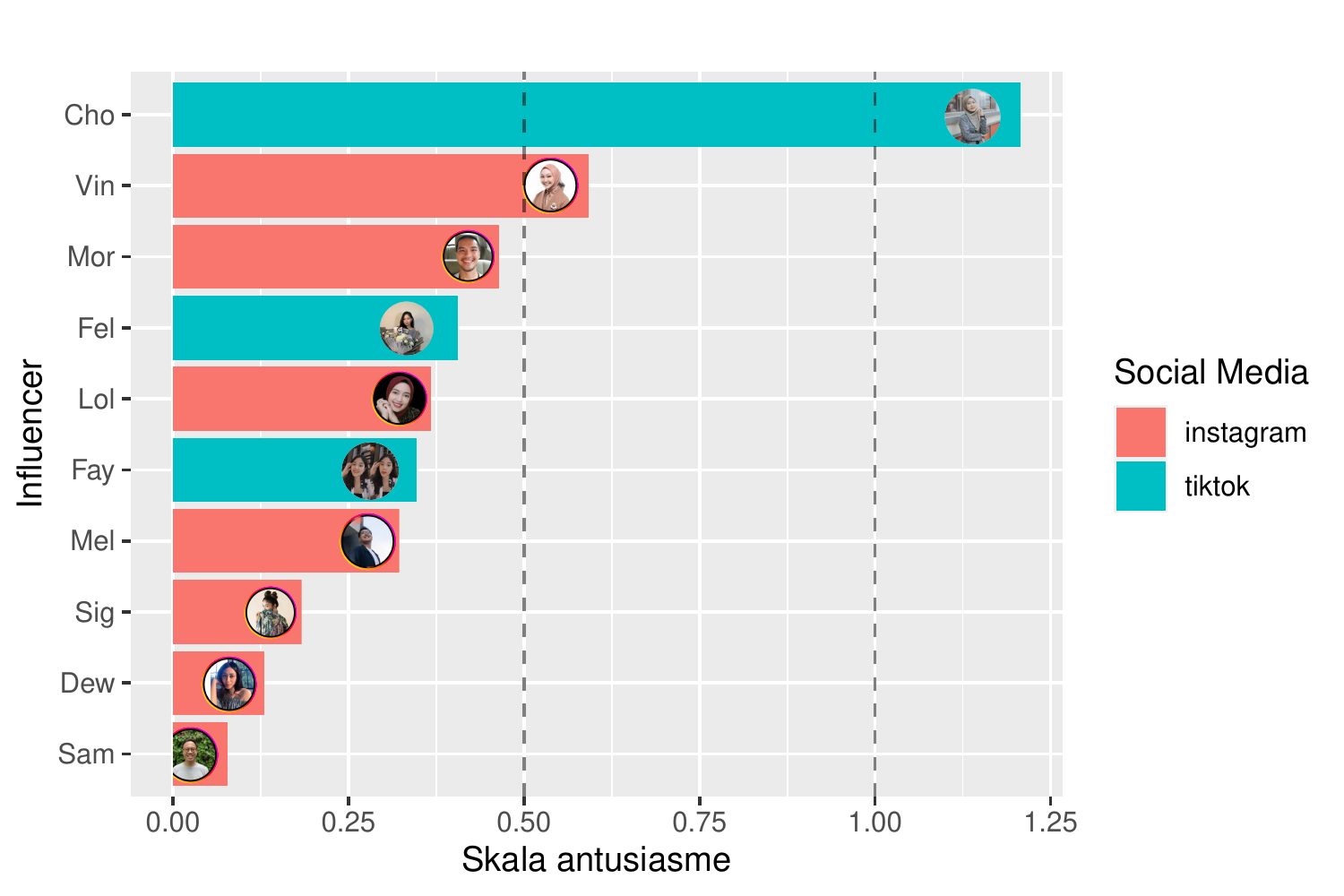} 

}

\caption{Tingkat Antusiasme Audiens per KOL}\label{fig:folent}
\end{figure}

Apabila ditinjau dari segi antusiasme follower KOL dengan parameter pengukuran yaitu \textbf{banyaknya jumlah like konten SBR011 dibandingkan rata-rata jumlah like per post dari tiap KOL}, maka KOL yang memiliki tingkat antusiasme follower tertinggi adalah Chornella Tri Pratama, dimana pengikut \emph{influencer} tersebut memberikan like pada konten SBR yang diunggah hampir mencapai \textbf{1,25 kali lipat} lebih banyak (125\%) daripada rata-rata like yang biasa diperoleh oleh influencer tersebut (grafik \ref{fig:folent}). Dengan kata lain, audiens dari Chornella adalah yang paling terkesan dengan iklan produk investasi SBR011.

Sementara itu, pengikut Vina Muliana terlihat kurang begitu tertarik dengan konten SBR011 yang diunggah terlihat dari penurunan jumlah like yang diterima ketika mengunggah konten SBR011 sebesar hampir 40\% lebih rendah dari rata-rata jumlah like yang biasa diterima. Pun demikian dengan KOL lainnya yang beberapa diantaranya memfokuskan unggahan konten pada tema investasi/finansial. Tampaknya produk SBR011 kurang menarik bagi audiens yang sebenarnya memiliki minat pada tema spesifik tersebut.

Dalam blog pribadinya (\url{https://desty.page/chornellatp}), Chornella menjelaskan profilnya sebagai seorang buruh sekaligus investor generasi Z \autocite*{Chor}. Status pekerjaannya sebagai buruh pabrik ini yang cukup membedakan sosok ini dari dua KOL lain yang sama-sama menggunakan TikTok pada kampanye SBR011 yaitu Felicia dan Fayza.

Dengan asumsi bahwa pengikut Chornella memiliki jenis pekerjaan yang serupa serta melihat tingginya antusiasme mereka terhadap unggahan konten SBR011, pemerintah sejatinya dapat mulai menjajaki peluang untuk menawarkan produk investasi retailnya kepada segmen pekerja kelas buruh terutama yang berusia muda, dimana sependek pengetahuan penulis jenis profesi \emph{blue collar} tersebut belum pernah mendapatkan perhatian khusus dalam kampanye kegiatan pemasaran SBN Ritel baik konvensional maupun yang berbasis syariah.

Berdasarkan publikasi BPS tentang Keadaan Angkatan Kerja di Indonesia Februari 2022, jumlah angkatan muda (usia 15-24 tahun) yang bekerja sebagai buruh/karyawan berjumlah sekitar 8,9 juta orang \autocite{BPS}. Angka tersebut tentu menunjukkan bahwa kategori pekerjaan tersebut merupakan pasar yang cukup potensial dan secara profil usia termasuk target utama penjualan SBN Ritel yakni investor berusia muda dan telah memiliki Kartu Tanda Penduduk.

\begin{figure}[H]

{\centering \includegraphics{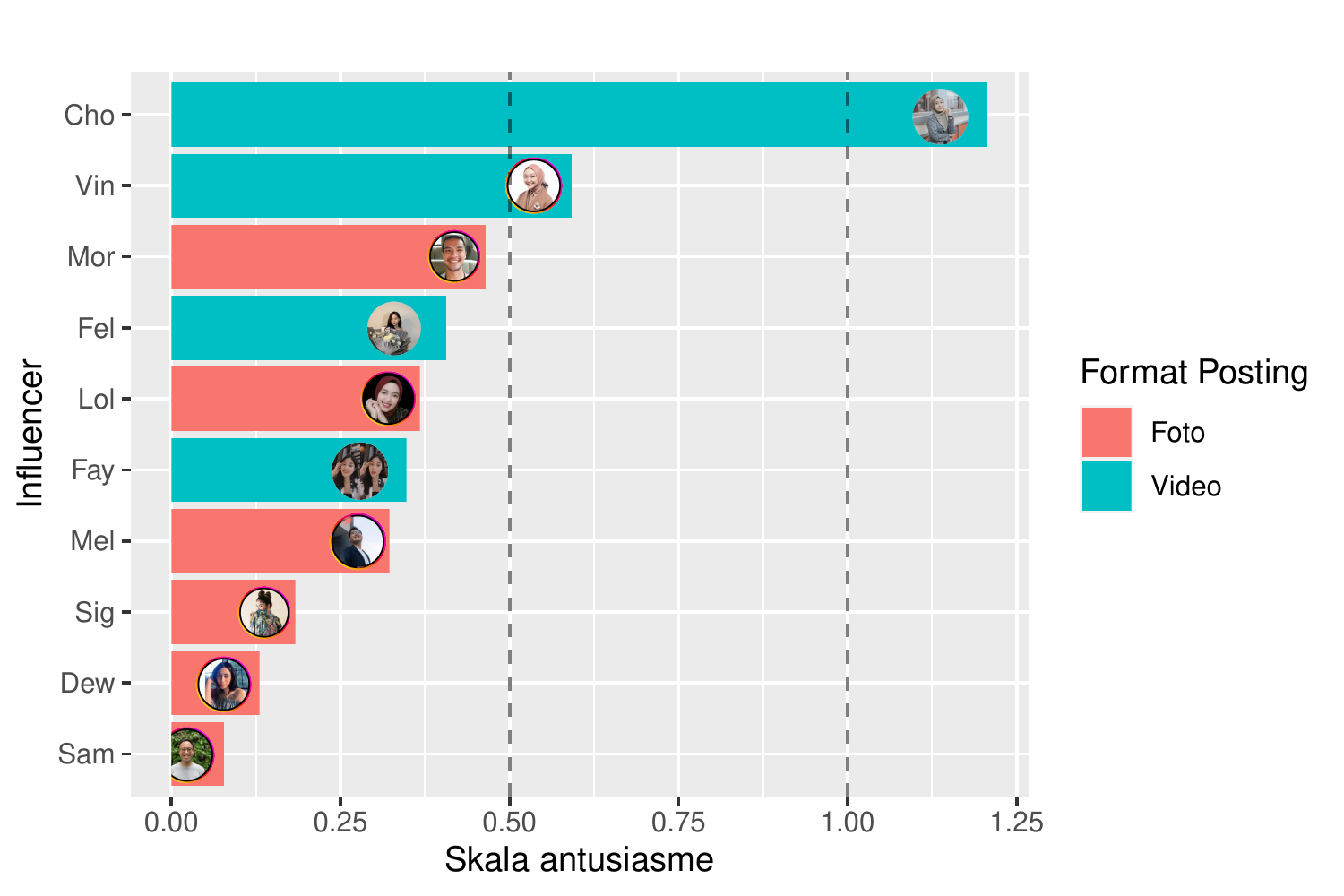} 

}

\caption{Tingkat Antusiasme Audiens berdasarkan Format Konten}\label{fig:format}
\end{figure}

Hal lain yang menarik dicermati adalah terkait format dari unggahan konten SBR011. KOL yang mengunggah \emph{campaign} dalam format video cenderung memiliki tingkat antusiasme yang tinggi dibandingkan dengan KOL yang menggunggah foto statis. Hal ini dapat dilihat pada grafik \ref{fig:format} di bawah ini.

\hypertarget{analisis-sentimen-audiens-terhadap-konten-sbr011}{%
\subsubsection{Analisis Sentimen Audiens terhadap Konten SBR011}\label{analisis-sentimen-audiens-terhadap-konten-sbr011}}

Untuk mendapatkan gambaran yang lebih menyeluruh mengenai penggunaan KOL pada pemasaran SBR011, selain dilakukan analisis terhadap tingkat engagement dan antusiasme audiens perlu dilakukan juga analisis sentimen terhadap \emph{feedback} yang diberikan oleh audiens tersebut.

Tahap awal yang dilakukan dalam proses ini adalah mendapatkan dataset komentar \emph{followers} dari KOL SBR011. Untuk memperoleh dataset dimaksud penulis menggunakan platform apify.com yang melakukan \emph{scrapping} terhadap seluruh komentar pada unggahan konten SBR011. Terdapat keterbatasan pada prosedur perolehan dataset ini dikarenakan komentar yang diambil hanya berasal dari KOL yang menggunakan media sosial Instagram. Adapun komentar yang berasal dari platform TikTok tidak dapat diperoleh karena aturan privasi data yang lebih ketat.

Dari beberapa KOL yang menggunakan media sosial Instagram, diperoleh 3 nama yang memiliki komentar \emph{followers} yang cukup banyak sehingga dapat digunakan untuk keperluan analisis sentimen. Ketiga nama tersebut adalah Vina Muliana, Morgan Oey dan Sigi Wimala. Dari beberapa komentar yang muncul, tidak semua mengomentari hal terkait investasi atau SBR011. Namun demikian, kata-kata tersebut tetap dimasukkan ke dalam dataset yang akan dianalisis lebih lanjut untuk melihat gambaran sentimen secara umum.

Proses penyiapan dataset kemudian dilanjutkan dengan kegiatan pembersihan (\emph{cleansing}) data dengan menyaring karakter selain huruf (\emph{string}), diantaranya angka, simbol serta tanda baca. Selanjutnya, dilakukan penghapusan kata-kata yang tidak membentuk makna bahasa natural, seperti kata sambung dan tunjuk menggunakan daftar \emph{stopwords} bahasa Indonesia dan bahasa Inggris yang terdapat dalam database NLTK dan ISO pada library Stopwords \autocite{stop}. Stopwords bahasa Inggris digunakan sebab beberapa komentar tersebut menggunakan kata dalam bahasa Inggris.

Tahap selanjutnya dalam proses analisis sentimen ini adalah melakukan ``stemming'' atau ``lemmatization'', yaitu mengubah suatu kata menjadi bentuk dasarnya untuk memudahkan dalam menganalisis kata yang unik. Daftar 10 kata unik yang paling sering muncul dalam komentar audiens KOL SBR011 dapat dilihat pada tabel \ref{tab:words}. Pengubahan kata ini menggunakan model training bahasa Indonesia dari library udpipe \autocite{udpipe}. Setelah dilakukan proses stemming, langkah berikutnya adalah memastikan bahwa kata unik yang muncul tersebut telah berformat baku. Proses ini memerlukan adanya penyesuaian terhadap kata-kata yang tidak baku, semisal mengubah nyidam menjadi mengidam dan ngelamar menjadi melamar.

\begin{table}[H]

\caption{\label{tab:words}Daftar 10 Kata Unik Terbanyak}
\centering
\fontsize{10}{12}\selectfont
\begin{tabular}[t]{l|r}
\hline
\cellcolor{green}{text} & \cellcolor{green}{n}\\
\hline
investasi & 5\\
\hline
banget & 4\\
\hline
barang & 4\\
\hline
mengatur & 3\\
\hline
uang & 3\\
\hline
bahan & 2\\
\hline
bank & 2\\
\hline
duit & 2\\
\hline
gudang & 2\\
\hline
juta & 2\\
\hline
\end{tabular}
\end{table}

Setelah memastikan bahwa setiap kata sudah berbentuk baku, kemudian kumpulan kata baku tersebut dianalisis kembali untuk memastikan tidak terdapat kata yang kurang bermakna. Dataset yang telah melalui proses tersebut dibawa ke dalam tahapan analisis selanjutnya yaitu pemberian skoring sentimen.

\begin{figure}[H]

{\centering \includegraphics[width=1\linewidth]{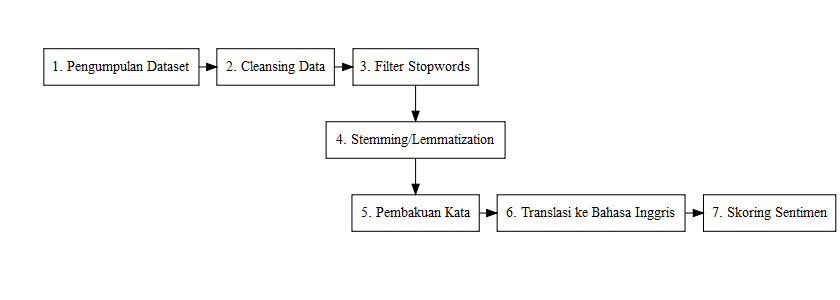} 

}

\caption{Flow Diagram Analisis Sentimen Audiens SBR011}\label{fig:flow}
\end{figure}

Skoring dilakukan terhadap setiap kata pada dataset dengan memanfaatkan katalog sentimen NRC \autocite{saif} didalam library syuzhet \autocite{syu}, yang membagi kategori sentimen menjadi 10 tipe yaitu; \textbf{anger, anticipation, disgust, fear, joy, negative, positive, sadness, surprise dan trust}. Dikarenakan library NRC hanya tersedia untuk kata dalam bahasa Inggris, perlu dilakukan translasi kata dalam dataset dari bahasa Indonesia menjadi bahasa Inggris. Proses alih bahasa dataset tersebut menggunakan library googleLanguageR \autocite{goo}. Hasil terjemahan kembali direviu untuk memastikan tidak terjadi perubahan makna dari kata aslinya di bahasa Indonesia. Bagan alur proses sentimen analisis secara lengkap sebagaimana ditunjukkan pada gambar \ref{fig:flow}

\begin{table}[H]

\caption{\label{tab:skor}Skoring Sentimen per Kata (Tidak Semua Kategori Ditampilkan)}
\centering
\fontsize{10}{12}\selectfont
\begin{tabular}[t]{l|l|r|r|r|r|r|r|r}
\hline
\cellcolor{green}{text} & \cellcolor{green}{translated} & \cellcolor{green}{anger} & \cellcolor{green}{anticipation} & \cellcolor{green}{disgust} & \cellcolor{green}{fear} & \cellcolor{green}{joy} & \cellcolor{green}{sadness} & \cellcolor{green}{surprise}\\
\hline
akal & sense & 0 & 0 & 0 & 0 & 0 & 0 & 0\\
\hline
allah & god & 0 & 1 & 0 & 1 & 1 & 0 & 0\\
\hline
ayo & come on & 0 & 0 & 0 & 0 & 0 & 0 & 0\\
\hline
bahan & ingredient & 0 & 0 & 0 & 0 & 0 & 0 & 0\\
\hline
banget & very & 0 & 0 & 0 & 0 & 0 & 0 & 0\\
\hline
bank & bank & 0 & 0 & 0 & 0 & 0 & 0 & 0\\
\hline
bantuan & help & 0 & 0 & 0 & 0 & 0 & 0 & 0\\
\hline
barang & goods & 0 & 0 & 0 & 0 & 0 & 0 & 0\\
\hline
batasan & limitation & 0 & 0 & 0 & 0 & 0 & 0 & 0\\
\hline
bca & bca & 0 & 0 & 0 & 0 & 0 & 0 & 0\\
\hline
\end{tabular}
\end{table}

\begin{figure}[H]

{\centering \includegraphics[width=0.9\linewidth]{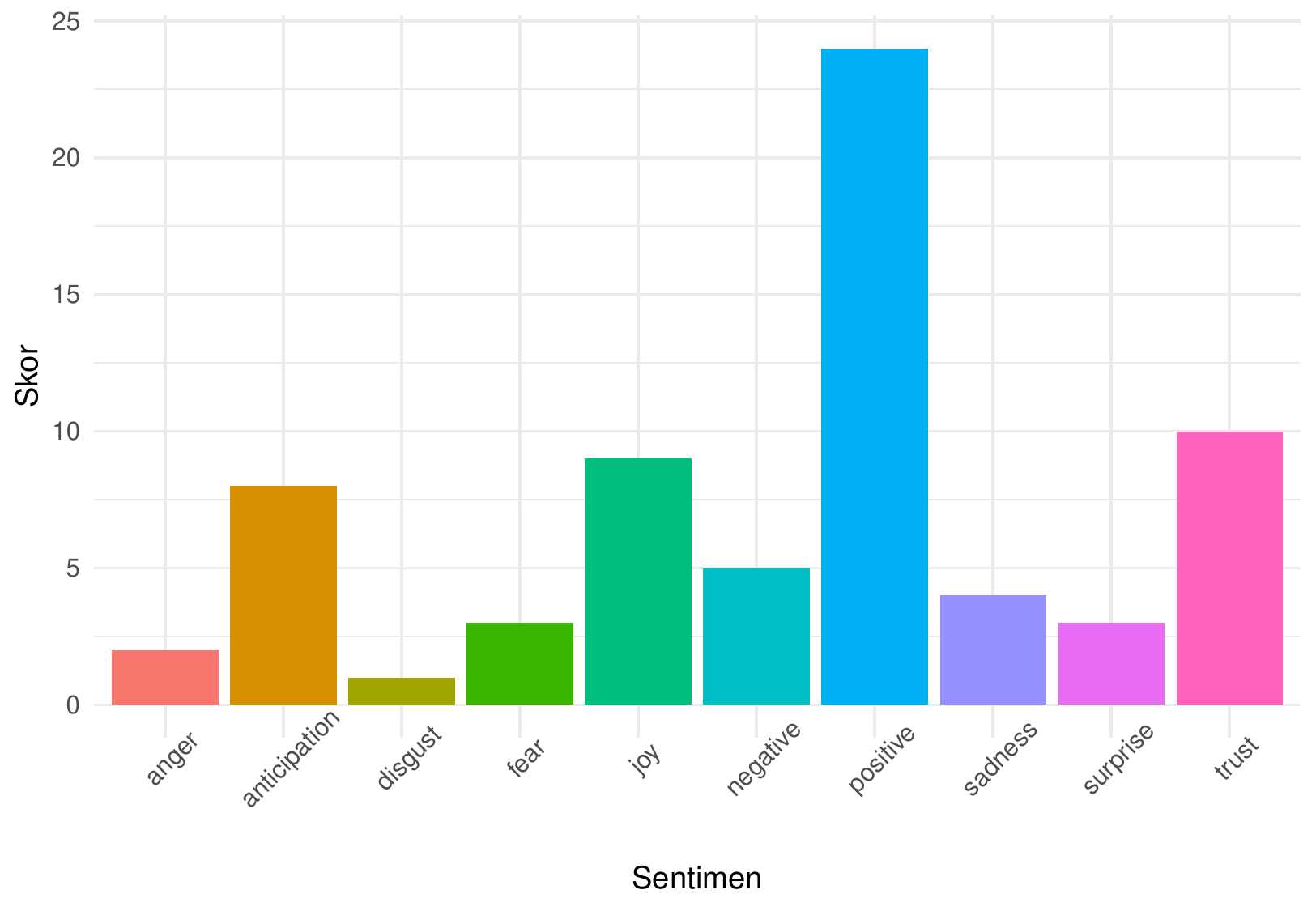} 

}

\caption{Sentimen Analisis dari Komentar Audiens KOL SBR011 yang Menggunakan Instagram}\label{fig:sentimen}
\end{figure}

Kata-kata yang sudah dalam bentuk bahasa Inggris tersebut pada akhirnya dapat dikelompokkan per kategori sentimen sebagaimana tampak pada tabel \ref{fig:sentimen}. Satu kata dapat memiliki beberapa kategori sentimen. Masing-masing kategori sentimen bernilai 1. Adapun total skor per kategori sentimen dapat dilihat pada gambar \ref{fig:sentimen}. \textbf{Dari hasil analisis sentimen diketahui bahwa secara umum mayoritas audiens memiliki respon positif (\emph{positive}) terhadap produk SBR011 yang dikampanyekan. Rasa kepercayaan (\emph{trust}), kegembiraan (\emph{joy}) dan antisipasi (\emph{anticipation}) juga turut mendominasi sentimen audiens pemasaran SBR011 oleh KOL.}

Kata-kata yang mengandung sentimen \emph{positive} diantaranya akal, berharga dan semangat. \emph{Trust} misalnya terdapat pada kata bank, kesaksian dan presiden, sementara kata-kata seperti cakep, uang dan gaji dikategorikan sebagai \emph{joy.} Adapun \emph{anticipation} terdiri dari kata-kata semisal Tuhan, uang dan membagikan.

\hypertarget{kesimpulan}{%
\subsection{Kesimpulan}\label{kesimpulan}}

\hypertarget{efektivitas-penggunaan-kol-pada-penjualan-sbr011}{%
\subsubsection{Efektivitas Penggunaan KOL pada Penjualan SBR011}\label{efektivitas-penggunaan-kol-pada-penjualan-sbr011}}

Meskipun nilai realisasi penjualan yang dihubungkan secara langsung dengan endorsement KOL SBR011 tidak dapat diketahui, tingkat engagement dan antusiasme audiens bisa menjadi penunjuk/\emph{proxy} terhadap keefektivitasan program tersebut. Secara umum, tampak bahwa \emph{engagement} audiens hanya berfokus kepada figur yang memang merupakan \emph{top influencer} seperti Vina Muliana (Forbes 30 Under 30). Namun demikian, beberapa sosok yang cukup tenar dan dikenal masyarakat luas semisal Morgan Oey terlihat belum dapat memberikan kontribusi yang signifikan dan sebesar apa yang dilakukan oleh Vina Muliana. Hal ini bisa berarti dua hal:\\
1. KOL yang populer belum tentu memiliki audiens yang memiliki minat terhadap dunia investasi;\\
2. KOL yang memiliki audiens dengan ketertarikan terhadap dunia investasi, belum tentu dapat mengarahkan perhatian audiensnya terhadap SBR011.

Berdasarkan analisis PCA terkait pengelompokkan KOL sebagaimana gambar \ref{fig:pca}, Vina Muliana tampak memiliki kombinasi fitur yang membuatnya berbeda dari kelompok KOL lain yakni popularitas dan kemampuan untuk membuat konten edukasi yang menyedot perhatian dan disukai audiensnya. Hal ini sejalan dengan hasil penelitian Wijaya dan Sugiharto \autocite*{wijaya} bahwa faktor \emph{power} dan \emph{atractiveness} dari selebritas yang di-\emph{endorse} signifikan dalam mempengaruhi \emph{engagement} produk.

Meskipun demikian, jika ditinjau dari rata-rata tingkat \emph{engagement} yang biasa diperoleh oleh masing-masing KOL terlihat bahwa \emph{engagement} konten tentang SBR011 yang dibuat oleh Vina mengalami penurunan sekitar 40\% dari rata-rata engagement yang diterima KOL tersebut. Hal itu menunjukkan bahwa tingkat antusiasme audiens Vina terhadap konten SBR011 secara umum jauh lebih rendah dibandingkan dengan konten-konten lain yang diunggahnya.

Hal berbeda ditunjukkan oleh audiens dari KOL Chornella T.P. Meskipun tingkat \emph{engagement} tidak sebesar pada KOL Vina Muliana, namun antusiasme audiens Chornella terhadap konten SBR011 jauh lebih tinggi bila dibandingkan rata-rata antusias mereka terhadap konten yang diunggah \emph{influencer} yang juga pekerja pabrik tersebut.

Apabila ditinjau dari format konten, konten SBR011 yang diunggah dalam format video lebih cenderung mampu menarik perhatian dibandingkan dengan konten berbentuk foto. Hal tersebut sebagaimana tercermin pada grafik \ref{fig:format}.

Selanjutnya, berdasarkan sentimen \emph{feedback} audiens terhadap kampanye SBR011, dapat disimpulkan bahwa secara umum kegiatan pemasaran yang dilakukan oleh KOL telah cukup baik dengan adanya sentimen \emph{positif, trust} dan \emph{joy} yang mendominasi sentimen audiens.

\hypertarget{implikasi-terhadap-kebijakan-dan-limitasi}{%
\subsubsection{Implikasi terhadap Kebijakan dan Limitasi}\label{implikasi-terhadap-kebijakan-dan-limitasi}}

Berdasarkan analisis yang dilakukan terdapat setidaknya dua hal yang dapat dipertimbangkan dalam pengambilan kebijakan kampanye pemasaran SBR011.\\
Pertama, antusiasme yang tinggi dari audiens KOL muda Chornella T.P. memberikan indikasi awal bahwa kemungkinan terdapat ceruk target segment yang dapat disasar oleh Pemerintah untuk kampanye penjualan SBR011. Mengingat bahwa KOL ini mengidentifikasi dirinya sebagai pekerja buruh sekaligus investor. Dengan asumsi bahwa audiensnya merupakan masyarakat dengan karakteristik serupa, Pemerintah dapat mempertimbangkan untuk menjangkau potensial investor muda dari kalangan \emph{blue collar}.\\
Kedua, melihat tingkat antusiasme audiens KOL yang mengunggah konten marketing dalam bentuk video (Vina, Chornella, Fayza dan Felicia) cenderung lebih baik ketimbang KOL yang menggunggah foto, pembuatan konten yang diunggah oleh KOL dapat diarahkan dalam format video alih-alih konten foto agar dapat menghasilkan engagement yang tinggi.

Tinjauan dalam paper ini memiliki keterbatasan yaitu hanya melihat keefektivitasan KOL dalam memasarkan penjualan SBR011 melalui indikator tidak langsung (\emph{proxy}) meliputi tingkat \emph{engagement}, antusiasme dan sentimen audiens. Untuk melihat dampak keuangan (\emph{return on investment}) dari penggunaan KOL terhadap penjualan SBR011, pemerintah maupun Mitra Distribusi kiranya perlu mempertimbangkan untuk menyediakan kode referensi/\emph{referral} yang dapat diinput oleh calon investor pada saat pemesanan, sehingga dapat diperoleh informasi mengenai sumber KOL yang menggerakkan pemesanan tersebut.

\printbibliography

\end{document}